\documentclass %[referee]
{aa}

\usepackage{graphicx}

\begin{document}

\title{An unusually low mass of some ``neutron" stars?}

\author{D. Gondek-Rosi\'nska\inst{1,2}
\and W. Klu\'zniak\inst{2,3,4}
\and N. Stergioulas\inst{5}}

\institute{LUTH,
FRE 2462 du C.N.R.S., Observatoire de Paris, F-92195 Meudon Cedex, France
\and Copernicus Astronomical Center, ul. Bartycka 18, 00-716 Warszawa, Poland
\and Institute of Astronomy, University of Zielona G\'ora, ul. Lubuska 2,
Zielona G\'ora, Poland
\and 
 Institut d'Astrophysique de Paris, 98bis Boulevard Arago, 75014 Paris
France
\and Department of Physics, Aristotle University of Thessaloniki,
Thessaloniki 54124, Greece}

\offprints {D. Gondek-Rosi\'nska, \email{Dorota.Gondek@obspm.fr}} 

\date{Received, Accepted}

\abstract{
The X-ray emission of RXJ1856.5-3754
has been found to coincide to unprecedented accuracy
with that of a blackbody, of  radius $ 5.8\pm0.9$ km 
for the measured
parallax distance of $ 140\,$pc (Burwitz et al. 2001, Drake et al. 2002).
 If the emission is uniform over the whole surface
of a non-rotating star, the mass of the star cannot exceed
$0.75\pm0.12M_\odot$ regardless of its composition.
If the compact object is a quark star described by the MIT-bag equation of
state (a ``strange star''), the mass is no more than $0.3M_\odot$.
Comparably small masses are also obtained 
for the X-ray bursters Aql X-1 and KS1731-260 for some fits to their
spectra.
\keywords{dense matter - equation of state - stars: neutron - stars:
general - X-rays: stars}}

\maketitle
\section{Introduction}

As noted by several authors, conventional neutron stars always
have a (circumferential) radius larger than 6 km.  Recent reports of a
rather small blackbody radius of a nearby neutron-star candidate
have generated speculation that the compact
object may not be a neutron star but a quark star instead.
Here, we point out that
although the actual composition of stars with a 6 km
radius is not known, what would make such stars
unusual is their low mass, posing a challenge to current theories of
their formation.
% of such objects.
 Detailed simulations of supernovae
do not predict remnant masses
less than $ ~1.2 M_\odot$ (Timmes et al. 1996).

\section{Dense matter and compact objects}

The properties of bulk matter at about nuclear density are
not well understood. On one hypothesis, its stable form
is composed of deconfined up, down and strange quarks
in about equal numbers, and quark stars should exist
(Itoh 1970, Bodmer 1971, Witten 1984). On another,
the lowest energy state of matter at supranuclear density
consists mainly of neutrons. Some observations of young
pulsars, specifically of impulsive changes (glitches) in the
radio period, seem to favor the latter hypothesis (Alpar 1987).

Much effort has been expended in trying to constrain
the equation of state (e.o.s.) of very dense matter
through the comparison of calculated and observed properties
of neutron stars. Unfortunately, the density and angular momentum
of the observed ``neutron'' stars are poorly constrained.
The rotational periods of radio pulsars have been measured
with exquisite accuracy, but the masses remain largely unknown.
In the few cases where the masses have been measured, there is
hardly any information on the radius. For the accreting
``neutron'' stars in persistent low mass X-ray binaries
some idea about the radius
can be gleaned from spectral data, but the masses are very
uncertain (although said to be consistent with $\sim1.4M_\odot$),
 the rotational periods also remain largely unknown.

Once the e.o.s. is selected, calculating the structure and space-time
metric of a compact object presents no fundamental difficulty.
Detailed numerical models of neutron stars have been computed
for a range of conventional e.o.s. of baryonic matter
(Arnett and Bowers 1977, Cook et al. 1994, Lattimer \& Prakash 2001).
Ditto for quark stars (Alcock et al. 1986, Haensel et al. 1986,
Gourgoulhon et al. 1999, Stergioulas et al. 1999, Gondek-Rosi\'nska et al.
2000, 2001). One essential difference between
conventional  neutron stars and quark stars
is that if quark matter is stable, there is no lower limit to the
mass of quark stars. Fully relativistic numerical computations
show that at masses below  $\sim 0.1M_\odot$
rotating quark stars are very well approximated by Maclaurin spheroids
(Amsterdamski et al. 2002). The maximum mass of quark stars
falls in the conventional range of maximum neutron star masses,
it does not exceed $\sim2.6 M_\odot$ for static models (Zdunik~et~al.~2000)
and $\sim3.7 M_\odot$ for rapidly rotating models (Stergioulas et al. 1999),
at least in the MIT-bag model of quark matter.

An unconventional e.o.s. of baryonic matter has also been proposed,
which could have densities below or above nuclear, depending on the choice
of parameters---Bahcall, Lynn, and Selipsky (1989) show, using an
effective field theory approach, that self-bound bulk
baryonic matter is consistent with nuclear physics data and low-energy
strong interaction data.
For neutron stars modeled with this e.o.s.
very low masses  are allowed, as for quark stars,
while the maximum mass depends on the choice of parameters
and could be lower or much higher than that of the Hulse-Taylor
binary pulsar ($1.4M_\odot$), these are the so called Q-stars
(Bahcall et al. 1990).
We stress that Q-stars would be composed  of 
hadronic matter, baryons and mesons. This matter differs from the one
considered in conventional neutron star models
only in the detailed description of
nuclear interactions. 

Fig.~1 illustrates some models of compact stars
computed by solving the TOV equation (Oppenheimer and Volkoff 1939)
with the equation of state 
$$P=a(\rho-\rho_0)c^2. \eqno (1)$$
 Here, $P$ is the pressure,
$\rho c^2$ the energy density, $\rho_0$ the density at zero pressure,
and $a$ a parameter.
In the MIT-bag model of quark matter (Farhi and Jaffe 1984),
quark confinement is modeled with a non-zero
 energy density of the vacuum, e.g.,  $B=\rho_0c^2/4$ when the quarks
are massless and $a=1/3$ is also obtained.
Zdunik (2000) gives expressions
for $a$ and $\rho_0$ as  functions of the quark masses and the QCD coupling
constant. With
a different model of quark confinement, Dey et al. (1998) derive an e.o.s.
which to a very good approximation is the same as eq.~(1)
(Gondek-Rosi\'nska et al. 2000).
For illustrative purposes
we reproduce one of the  Dey et al. (1998) sequences of models, 
the one with a maximum gravitational mass of $1.44 M_\odot$,
for which $a=0.463$ and $\rho_0=1.153\times10^{15}{\rm g\ cm^{-3}}$.

To compute  sequences of mainstream (MIT-bag) models of quark
stars, we have used
${\rm m_s=200\, MeV}$ for the mass
of the strange quark, a value of
$\alpha=0.2$  for the QCD coupling constant
and a bag constant $B=56\, {\rm MeV/fm^3}$, corresponding to 
$ \rho_0=4.50\times10^{14}\,{\rm g\cdot cm^{-3}}$, and $a=0.301$ in eq. (1), 
for the ``MIT SS1'' curve; and $m_{\rm s}=100\ {\rm MeV}$,
$\alpha=0.6$, $B=40\ {\rm MeV/fm^3}$,
 corresponding to $a=0.324$, $\rho_0=3.056\times10^{14}\,{\rm g\cdot cm^{-3}}$
 for the ``MIT SS2'' curve. 
These sequences allow gravitational masses of quark stars to be as
high as any value reported for the observed ``neutron'' stars
(within error bars).
The models were computed with and without a crust. For the crust
we use the BPS e.o.s. (Baym et al. 1971). 
The maximum density of the crust is taken to be
equal to the neutron-drip density 
$\rho_{\rm drip}=4.3 \times10^{11}{\rm g\ cm^{-3}}$, but thinner crusts
are not excluded.

For the Q-stars, we have chosen parameters in such a way that
the e.o.s. formally coincides with that of eq.~(1), and
the maximum static mass is about $1.0M_\odot$. With this choice for
the three sequences of stellar models,
at any given stellar mass, the Q-star, which is a neutron star 
really (i.e., a star composed
of baryonic matter), is the one with the smallest radius,
and the MIT-bag quark star is the least compact.
With another choice of parameters, the Q-star would have the largest
radius.
All three types of stars considered are typically more compact
than conventional neutron stars.

\section{The blackbody radius in the Schwarzschild metric}

In this letter we neglect rotation of the star.
The exterior of any spherically symmetric star is described by
the Schwarzschild metric. We also assume that the stellar radius
satisfies $R>3GM/c^2$, this is true for all the models considered
in Fig.~1.

If the stellar surface radiates as a blackbody,
the spectrum and luminosity at infinity satisfy
the Stefan-Boltzmann law with a blackbody radius 
$$R_{\rm bb}=R/\sqrt{1-2GM/(Rc^2)},\eqno (2)$$ 
where $R$ is the
circumferential radius of the star and $M$ its mass.
The same formula applies when $R_{\rm bb}$ is determined from an
effective temperature derived by fitting the spectra to theoretical models of
neutron-star atmospheres (e.g., Rutledge et al. 2001a).
All recent conventional neutron-star models have the property that
$R_{\rm bb}>12 \,{\rm km}$
 (Lattimer \& Prakash 2001, Haensel 2001).
A star with a blackbody radius significantly smaller than 10 km probably 
cannot be described by a conventional e.o.s.
 
When eq. (2) is inverted
at any value of the blackbody radius, $M$
reaches a maximum at the photon orbit $R=3GM/c^2$.
Accordingly (Lattimer \& Prakash 2001), 
$$M<0.13M_\odot {R_{\rm bb}\over1.0\,{\rm km}}. \eqno (3)$$
Thus, a non-rotating star radiating as a blackbody cannot have a mass
greater than $1 M_\odot$ if the blackbody radius is 
$R_{\rm bb}<7\,{\rm km}$.  
This value is much less than the $1.4M_\odot$ masses
measured for the binary pulsars, assumed to be representative of the
neutron star mass at its birth in a supernova collapse.

The limit of eq. (3) is shown in Fig.~1 as the  dashed line. 
There are no solutions to the redshift eq. (2) for any value
of $R_{\rm bb}$ above this line---the excluded area is shown in gray.
Also shown are some representative models of ultra-compact
stars (Section 2), as well as best-fit values of $R_{\rm bb}$,
 reported in the literature under certain assumptions as to the
spectrum,  for  three acctually observed compact objects (Sections 4, 5).

%%%%%%%%%%%%%%%%%%%%%%%%%%%%%%%%%%%%%%%%%%%%%%%%%%
\begin{figure*}
\centering
\includegraphics[width=14cm]{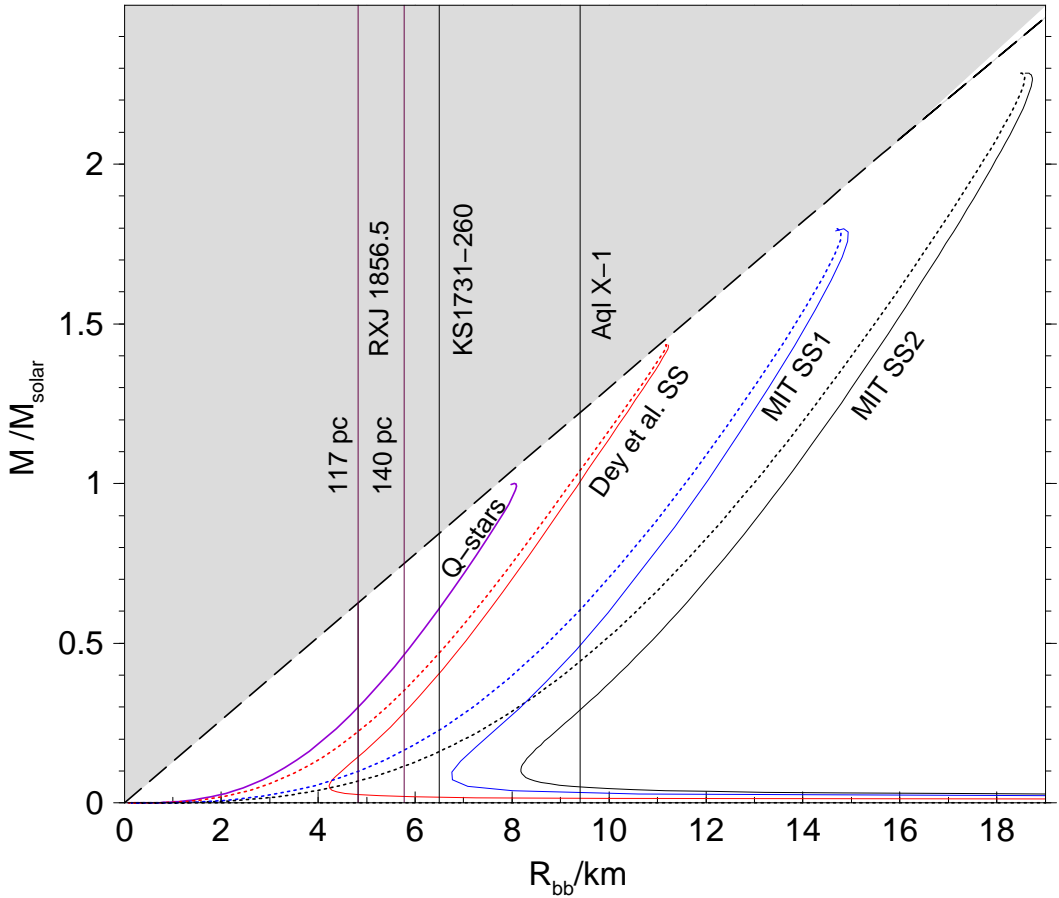}
\caption{The upper limit to mass allowed by eq. (2),
 and the masses of theoretical
models of compact objects, plotted as a function of the blackbody radius
in Schwarzschild geometry. The excluded area is shown in grey. 
The best fit values of the 
blackbody radius of RXJ 1856.5-3754 (Drake et al. 2002) are
shown as vertical lines for two values of the distance to the source,
$d=117\,$pc and $140\,$pc; the quoted $1\sigma$ error on $R_{\rm bb}$
is $0.68\,{\rm km}/100\,{\rm pc}$. Also shown are the best fit
radii for ``hydrogen''atmospheric models
of well-known X-ray bursters  KS1731-260 and Aql X-1 
(Rutledge et al. 2001a,b). The thin continuous lines are the
mass-radius relationship for three sequences of quark star models with 
the thickest possible crust of normal atomic matter,
the dotted lines are the same models without
the crust (``bare quark stars''), models with intermediate thickness
of the crust would fall in between these limiting lines.
 The thick continuous line is a
sequence of neutron star models (Q-stars) based on an unconventional
equation of state of baryonic matter (Bahcall et al. 1990).
See text for details.
}
\end{figure*}
%%%%%%%%%%%%%%%%%%%%%%%%%%%%%%%%%%%%%%%%%%%%%%%%%%%%%%%%%%%%%%%%%%

\section{RXJ1856.5-3754}

Shortly after its discovery with ROSAT,
Walter et al. (1996) and Neuh\"auser et al. (1997)
suggested that the steady, dim 
($\sim 1.5\times 10^{-11}\,{\rm erg\cdot s^{-1}\cdot cm^{-2}}$)
X-ray source RXJ1856.5-3754 is a nearby neutron star.
Pavlov et al. (1996) performed the first spectral fits
to the ROSAT data, and found strong dependence of the
results on chemical composition. 

The X-ray spectrum of the source observed with Chandra
is inconsistent with
that expected from a hydrogen, helium or iron
rich neutron star atmosphere, solar composition is also excluded
(Burwitz et al. 2001). 
The spectrum is fit to an extraordinary
accuracy by a blackbody, and the
best fit blackbody radius is 
$R_{\rm bb}=(4.12\pm0.68)\,{\rm km}\times d/(100\,{\rm pc})$
 (Burwitz et al. 2001, Drake et al. 2002); $d$ is the distance to the
source.

Drake et al. 2002 come to the conclusion that the new distance determinations,
especially the optical 
parallax determination of Kaplan et al.
(2002), firmly place RXJ1856.5-3754 at less than 140 pc distance.
On the tacit assumption that the whole surface of the star is emitting
uniformly, and that the star is not rotating rapidly,
Drake et al. further suggest that
the inferred radius is too small
for the star to be a neutron star, and the star may be a quark star
instead.
The upper limit to the pulsed fraction in the frequency range 
$10^{-4}\,$Hz to 100 Hz is less than 2.7\%.
The possibility that   RXJ1856.5-3754 is a millisecond pulsar
still remains.

A bare quark surface is a very weak photon emitter
(Chmaj et al. 1991, Usov 2001), and normal matter is usually expected
to have an atmosphere, so the observed spectrum  is a
major puzzle in itself. In this letter we focus on the radius alone.
Small blackbody radii have been found for the polar caps
of active pulsars, but in those cases a power-law component
of the X-ray spectrum and pulsations have also been detected.

If the X-ray source is indeed a star of blackbody
radius of about 6~km, or less,
it must be of unusually low mass for a compact stellar remnant,
 less than $0.8 M_\odot$ by eq. (3).
If it is a quark star of the usually considered properties, its mass
must be extremely small, $0.1M_\odot$ to within a factor of two
(as seen in Fig.~1 for the MIT-bag models).

\section{Are low mass ``neutron'' stars common?}

RXJ1856.5-3754 is only $10^2\,$pc away. On statistical grounds,
it cannot be an uncommon object. There should be about 
$10^5\cdot(10^{10}\,{\rm y}/\tau)$
such objects in the Galactic disk (assuming a half-thickness of 1 kpc),
where $\tau$ is the (cooling) time for the
RXJ source to become undetectable at 140 pc.
Are the inferred mass and radius of RXJ1856.5-3754 unique in the observed
``neutron'' star population? 

There are hints in the literature
that other stars of mass clearly less than the canonical $1.4M_\odot$
may have been observed. 
Steeghs and Casares (2002) have measured
the mass function of Sco X-1. A mass of $1.4M_\odot$
is consistent with the observed light curve, but
a somewhat lower value $\sim 0.9M_\odot$ is
more likely a priori.

Spectral fits to some X-ray bursters are especially interesting
in this regard. The derived mass and radius stongly depend
on the atmospheric model. Whether the blackbody
emission of  RXJ 1856.5-3754 originates from the whole surface of
the compact stellar source, or only a part of it, spectral
fits exclude conventional neutron-star atmospheres.
Is there a compelling reason  in other sources to prefer atmospheric
models not required by the data?

Aquila X-1 is a case in point. A hydrogen atmosphere without
a power law component fits the data for a radius of
$R_{\rm bb}=9.4^{+2.7}_{-2.4}\,$km, 
and with a power-law for a larger radius of
about $14\pm4\,$km (Rutledge et al. 2001a). It is the latter value
which is usually adopted, but in Fig.~1 we plot the former.
A blackbody fit yields the even smaller radius
of $R_{\rm bb}=1.9\pm0.3 \,$km.

A hydrogen atmosphere fit to another X-ray burster, KS1731-260,
yields a comparable radius $R_\infty=6.5^{+6}_{-3}\,$km
(Rutledge et al. 2001b).
We plot the central value in Fig.~1.
A blackbody fit gives $R_{\rm bb}=1.3^{+0.6}_{-0.3} \,$km.
The point we are making here is that for some X-ray
bursters observations do not exclude
low radii and low masses.

\section{Is RXJ1856.5-3754 a binary system?}

There is an excess optical emission  (over the extrapolated X-ray blackbody)
of  RXJ1856.5-3754, and it can also be fit by
a blackbody, with a lower limit to the blackbody
radius of $17\, {\rm km}$
(Burwitz et al. 2002).
The optical
source is constrained to be within $2''$ of the X-ray source.

If so,
we would like to suggest that a binary system is being observed,
one component being a larger (in radius) and cooler object,
while the other significantly smaller and hotter,
but both at about nuclear density 
(few times $10^{14}\,$ g cm$^{-3}$).
It is reasonable to assume that both members of the putative
pair have the same composition. The MIT-bag e.o.s. of quark matter
can accomodate these two disparate radii (less than 6 km and more than 17 km)
in two ways. The binary could be a massive ($\sim2M_\odot$) star with
a low mass satellite. Alternatively, both stars could be
of very low mass, and
differ in the thickness of the crust. As is clear from Fig.~1,
the maximal thickness of the crust increases with decreasing
 mass, the ``larger'' optical source
could then be of even lower mass than the X-ray source. Extremely low mass
quark stars with a thick crust are stable (Glendenning 1995, Gondek 1998).

Whether the two objects have been formed at the same time, perhaps as
a result of fragmentation of a rapidly rotating collapsing core,
or whether one is older and the other younger is at present a matter of
speculation, as no evolutionary calculations producing such a pair
have been performed. However, we note that Newtonian simulations
indicate that under violent circumstances quark stars are subject to
fragmentation, in which low mass ($\sim 0.1M_\odot$) quark ``starlets"
may be formed (Lee and Klu\'zniak 2001).

\begin{acknowledgements} 
 It is a pleasure to thank Dr. J. Tr\"umper for valuable discussion.
 This work has been funded by the KBN grant
 5P03D01721, the Greek-Polish Joint Research and Technology Program
 EPAN-M.43/2013555, the EU Program ``Improving the Human Research
 Potential and the Socio-Economic Knowledge Base'' (Research Training
 Network Contract HPRN-CT-2000-00137), and by CNRS.
\end{acknowledgements}

\end{document}